\documentclass[twocolumn,showpacs,preprintnumbers,amsmath,amssymb]{revtex4}

\usepackage{graphicx}
\usepackage{dcolumn}
\usepackage{bm}

\begin{document}

\title{Flux-lattice melting  in  LaO$_{1-x}$F$_{x}$FeAs: first-principles
prediction}

\author{Jian-Ping Lv$^{1}$ and Qing-Hu Chen$^{2,1,^{\dag}}$
}
\address{
$^{1}$ Department of Physics, Zhejiang University, Hangzhou
310027, P. R. China\\
$^{2}$  Center for Statistical and Theoretical Condensed Matter
Physics, Zhejiang Normal University, Jinhua 321004, P. R. China }
\date{\today}

\begin{abstract}

We  report the  theoretical study of the flux-lattice melting in the
novel iron-based superconductor  $LaO_{0.9}F_{0.1}FeAs$ and
$LaO_{0.925}F_{0.075}FeAs$. Using the Hypernetted-Chain closure and
an efficient algorithm, we  calculate the two-dimensional
one-component plasma pair distribution functions, static structure
factors and  direct correlation functions at various temperatures.
The Hansen-Verlet freezing criterion is shown to be valid for
vortex-liquid freezing in type-II superconductors. Flux-lattice
meting lines for $LaO_{0.9}F_{0.1}FeAs$ and
$LaO_{0.925}F_{0.075}FeAs$ are predicted through the combination of
the density functional theory and the mean-field substrate approach.
\end{abstract}
\pacs{74.25.Qt, 74.25.-q, 74.70.-b,  05.70.Fh }

\maketitle

\section{Introduction}

It is well known that the practical application of superconductors
crucially depends on the high electricity transport without
dissipation, so the searching for new superconductors with a high
transition temperature has attracted considerable attention in
material science. Recently, the novel superconductivity in
iron-based layered superconductors has been reported, providing a
new path to the high temperature superconductivity. The $LaOFeAs$
under doping with $F^{-}$ ions at the $O^{2-}$ sites was first
discovered to exhibit superconductivity with $T_{c}=26K$ by
Kamihara et al \cite{kamihara}. Then the holes were introduced in
$LaOFeAs$ by partially substituting $La$ with $Sr$, the
superconductivity was observed  with $T_{c}=25$ K \cite{wen}.
Shortly after these studies, $T_{c}$ was pushed up surprisingly to
above 40 K when $La$ in $LaO_{1-x}F_{x}FeAs$ was substituted by
other rare earth elements \cite{chen,wang1,ren1}. In later
experiments, the maximum $T_{c}$ at about 55 K was achieved in
$SmO_{1-x}F_{x}FeAs$ \cite{ren3,liu} and $Gd_{1-x}Th_{x}OFeAs$
\cite{wang}. Though these superconductors have a high critical
temperature, the mechanism and the possible difference compared
with copper-based high-$T_{c}$ superconductors are still very
controversial. Obviously, more experimental and theoretical  work
are needed to understand the type-II superconductivity as well as
the microscopic mechanism in these  new iron-based
superconductors.

Since  the vortex-lattice solid (glass) state without linear
resistivity is crucial for the application of high-$T_{c}$
superconductors, the melting of the flux-lattice in bulk type-II
superconductors is of fundamental and practical significance
\cite{nelson,cubitt,brandt,safar,kwok,segupta,cd2,huchen,xh,cor,col}.
The elastic theory combined with the phenomenological Lindemann
criterion has been used to study the flux-lattice melting in layered
superconductors\cite{Houghton}, which is based on the instability of
the crystal. However, the regular flux-lattice structure loses
simultaneously in the melting, the information for vortex liquid is
then lacking, so the Lindemann criterion has drawback in the study
of the  vortex melting. The density functional theory has been
successfully used in the flux-lattice melting  of layered
superconductors\cite{segupta,cd2,col,cor,xh}. Due to the large
anisotropy in these superconductors, weak Josephson interaction
between layers can be neglected, both the interplane and intraplane
interaction between the pancake vortices are electromagnetic in
origin. The density functional method is usually well appropriate
for layered superconductors with large anisotropy. Recently, a
rather large anisotropy $\sim 15$ \cite{Singh} was found in
$LaO_{0.9}F_{0.1}FeAs$ and $LaO_{0.925}F_{0.075}FeAs$ (1111-type) by
a bond structure calculation. The strong anisotropy in these systems
was also confirmed recently in  experiments\cite{chenli}.

In this work, we study the flux-lattice melting in these two
samples in the framework of the density functional theory and the
mean-field substrate model proposed recently \cite{sub}.  In our
first-principles calculations, the input contains just  a few
parameters. The two-dimensional (2D) one-component plasma (OCP)
pair distribution functions (PDF), 2D OCP static structure factors
(SSF) and 2D  OCP direct correlation functions (DCF) are
calculated in the celebrated Hypernetted-Chain (HNC) closure. The
validity of the Hansen-Verlet freezing  criterion is analyzed. The
melting lines for  $LaO_{0.9}F_{0.1}FeAs$ and
$LaO_{0.925}F_{0.075}FeAs$  are plotted by using the mean-field
substrate approach.

The  paper is organized as follows. Section II describes the
calculation of 2D OCP PDF. In section III, we analyze the
Hansen-Verlet freezing criterion and present the calculation of
flux-lattice melting lines. A short summary is given in the last
section.

\section{Calculation of Pair Distribution Functions}

The electromagnetic vortex-vortex interaction is given by a
three-dimensional (3D) anisotropic pair potential. In Fourier
space, it reads \cite{segupta,cor}
\begin{equation}
\beta V(\mathbf{k})= \frac
{\Gamma\lambda^{2}_{ab}(T)(k_{\bot}^{2}+(4/d^{2})\sin^{2}(k_{z}d/2)}
{k_{\bot}^{2}[1+k_{\bot}^{2}\lambda^{2}_{ab}(T)+(4\lambda^{2}_{ab}(T)/d^{2})\sin^{2}(k_{z}d/2]},
\end{equation}
where $\Gamma=\beta d\phi_{0}^{2}/4\pi\lambda^{2}_{ab}(T)$ and
$\beta=1/k_{B}T$, $k_{\bot}$ and  $k_{z}$ denote the component of
$\mathbf{k}$ perpendicular and parallel to the $c$-axis
respectively,
$\lambda_{ab}(T)=\lambda_{ab}(T=0)/[1-(T/T_{c})^{4}]^{1/2}$ is the
planar penetration depth, $\phi_{0}=hc/2e$ is the flux quantum and
$d$ is the layer spacing between FeAs layers. In our calculation,
$d$ is twice lattice parameter $c$, i. e. $d \approx$ 17.5 \AA
\cite{kamihara,Singh}.  According to Eq. (1), the intralayer
potential is repulsive while the interlayer potential is
attractive. Some parameters  measured in a muon spin relaxation
study \cite{luetkens} are used in the present work , e. g.
$\lambda_{ab}(T=0)=2540$ \AA, $T_{c}=26$ K for
$LaO_{0.9}F_{0.1}FeAs$  and $\lambda_{ab}(T=0)=3640$ \AA,
$T_{c}=22$ K for $LaO_{0.925}F_{0.075}FeAs$. For convenience, the
mean inter-particle spacing is taken as the length unit in the
following.

In the classical liquid theory, PDF is defined as \cite{sl}
\begin{equation}
g(r,r^{'})=\frac{\rho(r,r^{'})}{\rho(r)\rho(r^{'})},
\end{equation}
where $\rho(r)$ represents the probability to find a particle at
$r$, $\rho(r,r^{'})$ is the probability to find two
particles at $r$ and $r^{'}$ respectively, and the pair
correlation function $h(r)$ is defined by $h(r)=g(r)-1$.

In the uniform and isotropic liquid, the Ornstein-Zernike relation
can be written as  \cite{sl}
\begin{equation}
h(\mathbf{r})=C(\mathbf{r})+\rho\int
d\mathbf{r^{'}}C(|\mathbf{r}-\mathbf{r}^{'}|)h(\mathbf{r}^{'}),
\end{equation}
while in the  Fourier space, it becomes
\begin{equation}
h(\mathbf{k})=C(\mathbf{k})+\rho C(\mathbf{k}) h(\mathbf{k}).
\end{equation}

\begin{figure}[tbp]
\centering
\includegraphics[width=7.5cm]{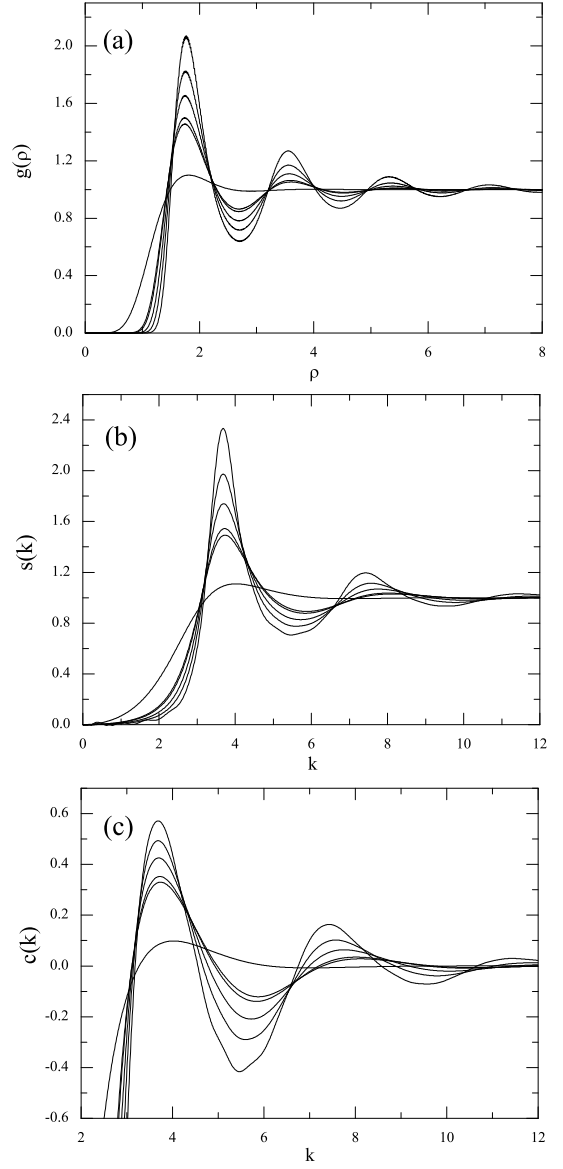}
\caption{\label{Figure Graph1} (a) 2D OCP PDF (b)  2D OCP SSF  and
(c)  2D OCP DCF for $LaO_{1-x}F_{x}FeAs$ (x=0.1,0.075) at
different temperatures. In all figures, from top to bottom at the
first peaks: 10K for $LaO_{0.9}F_{0.1}FeAs$, 14K for
$LaO_{0.9}F_{0.1}FeAs$, 10K for $LaO_{0.925}F_{0.075}FeAs$, 20K
for $LaO_{0.9}F_{0.1}FeAs$, 14K for $LaO_{0.925}F_{0.075}FeAs$,
and 20K for $LaO_{0.925}F_{0.075}FeAs$.}
\end{figure}

The HNC closure is widely used in the vortex system of type-II
superconductors \cite{segupta,cd2,cor}. In this closure, the
bridge function is set to zero and the DCF can be written as
\begin{equation}
C(r)=exp [ -\beta V(r)+Y(r) ]-Y(r)-1,
\end{equation}
where $Y(r)=g(r)-C(r)-1$. Following the method proposed in Ref.
\cite{segupta}, we extract  the correlation functions by splitting
$C(r)$ and $Y(r)$ into a short- and a long-ranged part as
\begin{equation}
 C^{n}(\rho)=C^{n}_{s}(\rho)+C^{n}_{l}(\rho)
\end{equation}
\begin{equation}
Y^{n}(\rho)=Y^{n}_{s}(\rho)+Y^{n}_{l}(\rho),
\end{equation}
here $Y^{n}_{l}(\rho)=\beta V(\rho,nd)$, $C^{n}_{l}(\rho)=-\beta
V(\rho,nd)$ and $C^{n}_{s}(\rho)=C^{0}_{s}(\rho)\delta_{n,0}$,
 $\rho$ denotes the in-plane coordinate and $n$ is the layer
index \cite{segupta}. In this way, the HNC equation in the short
range part can be written as
\begin{equation}
C^{0}_{s}(\rho)= exp [Y^{0}_{s}(\rho) ]-Y^{0}_{s}(\rho)-1.
\end{equation}
The in-plane PDF is then given by
\begin{eqnarray}
g^{0}(\rho) &=& C^{0}(\rho)+Y^{0}(\rho)-1\nonumber \\
&=&C^{0}_{s}(\rho)+Y^{0}_{s}(\rho)-1 \nonumber \\ &=&
exp(Y^{0}_{s}(\rho)).
\end{eqnarray}

Using an efficient algorithm proposed previously \cite{gillan}, we
are able to calculate the short range part of 2D DCF and $Y(\rho)$
successfully and then obtain  the in-plane PDF. These functions are
tabulated in a discrete form at about $1200$ points. Since the PDF
decreases rapidly in the order of the mean inter-particle spacing,
they can be  cut off in the real space to simplify the calculation.
The criterion for this cutoff is that  the converged functions are
not influenced. In our calculation, all of the converged functions
are stable and the  relative difference is less than $10^{-9}$.
Actually, the value of PDF saturates at high field,  consistent with
the well-known fact that the vortex system decouples to a 2D system
at high field. These PDF  can be regarded as the 2D OCP one, which
is shown in Fig. 1(a)  for $LaO_{1-x}F_{x}FeAs$ (x$=$0.1, 0.075). It
displays an oscillating behavior and becomes very small at large
$\rho$.

\begin{figure}[tbp]
\centering
\includegraphics[width=7cm]{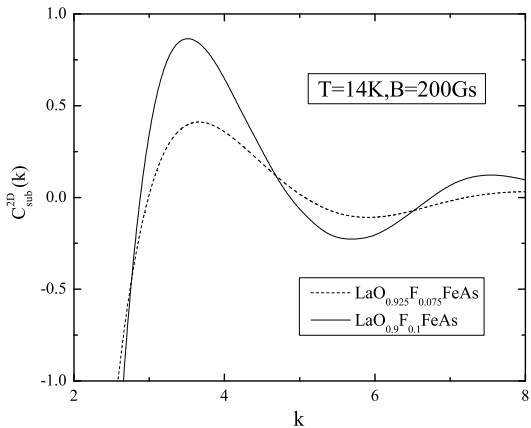}
\caption{\label{Figure Graph3} DCF based on the mean-field
substrate model for $LaO_{1-x}F_{x}FeAs$ (x=0.1,0.075) at T=14K,
B=200Gs.}
\end{figure}

\section{Freezing  criterion and  melting lines}

In the classical liquid theory, the SSF is defined as
\begin{equation}
S(\mathbf{\textbf{k}})=1+\rho\int
d\mathbf{r}(g(r)-1)exp(i\mathbf{k}\cdot\mathbf{r}).
\end{equation}
To provide  a visual representation of the vortex structure, we
calculate the 2D OCP SSF according to Eq. (10). The 2D OCP DCF are
then easily achieved with the relationship $S(k)=1/(1-C(k))$. Both
2D OCP SSF and 2D OCP DCF  are shown in Figs. 1(b) and (c). In fact,
the maximum values of these functions are  found to decrease as
temperature increases, shedding an insight of the 2D flux-lattice
system in the melting process. Furthermore, all these functions for
$LaO_{0.9}F_{0.1}FeAs$ exhibit a larger oscillation than those for
$LaO_{0.925}F_{0.075}FeAs$ at the same temperatures, indicating that
the 2D lattice structure in $LaO_{0.9}F_{0.1}FeAs$ is more clear
than that in $LaO_{0.925}F_{0.075}FeAs$.

In the density functional approach first suggested by Ramakrishnan
and Yussouff \cite{ry}, the free energy function is given by
\begin{eqnarray}
 \beta\triangle \Omega&=&\int d \mathbf
{r}[\rho(\mathbf{r}) ln\frac {p (\mathbf{r})}{\rho_{l}}-
\delta\rho(\mathbf{r})] \nonumber \\ &-&\frac{1} {2} \int dr \int
dr^ {'} C(|\mathbf{r}-\mathbf{r}^{'}|) \nonumber \\ &\times& [
\rho(\mathbf{r})-\overline{\rho}^{3D} ] [\rho
(\mathbf{r}^{'})-\overline{\rho}^{3D}],
\end{eqnarray}
here $\rho(\textbf{r})$ is the density at $\textbf{r}$,  $
\overline{\rho}^{3D}$ is the three-dimensional  homogenous liquid
density, and $C(|\mathbf{r}-\mathbf{r}^{'}|)$ is the two-body fluid
phase DCF. The two-body interactions between the particles are
expressed in the second integral on the right hand side.

\begin{figure}[tbp]
\centering
\includegraphics[width=8cm]{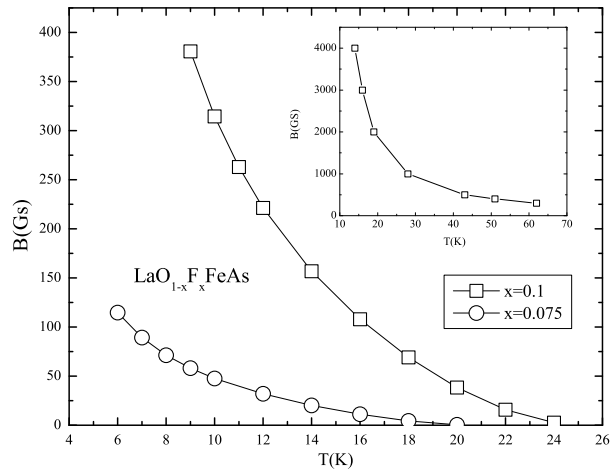}
\caption{\label{Figure Graph4} Flux-lattice melting lines for
$LaO_{1-x}F_{x}FeAs$. Inset: Flux-lattice melting line for $BSCCO$
calculated in Ref. \cite{cor}. }
\end{figure}

Recently, a mean-field substrate model was proposed to investigate
the thermodynamic properties in layered superconductors \cite{sub}.
Here we describe it briefly for completeness. Since the interlayer
interaction between vortices is much  weaker than the intralayer one
in the strongly anisotropic superconductors, the vortices can be
regarded as a collection of pancake vortices. In this way, the
interplane and out-of-plane interaction could be separated and the
DCF can be written as \cite{col}
\begin{equation}
C_{z}(\rho)=dC^{2D}(\rho)\delta(z)-\frac{V_{z}(\rho)}{T},
\end{equation}
where $ V_{z}(\rho)$ is the out-of-plane interaction potential. In
 Fourier space, the effective DCF of the mean-field substrate model takes
the following form

\begin{equation}
C^{2D}_{sub}(K) = C^{2D}(K)-\int dz \frac{\overline{\rho}
V_{z}(K)}{Td}.
\end{equation}

We employ the above equation to extract the DCF based on the
substrate model at different temperatures and magnetic fields. For
instance, we present  the DCF at T=14K and  B=200Gs in Fig. 2. As
proposed in Ref. \cite{col}, the single order parameter
$C^{2D}_{sub}(G)=0.856$
($(G=8\pi^{2}\overline{\rho}/\sqrt{3})^{1/2}$) can be used to
determine the flux-lattice melting lines, which is equivalent to
the nontrivial solution for $\beta\triangle \Omega=0$. On the
other hand, the Hansen-Verlet freezing criterion is successfully
used in the classical liquid theory \cite{hv}. This criterion
states that the liquid freezes when the first peak of the
structure factor reaches a critical value. The density functional
theory in the vortex system of type-II superconductors is also
based on the classical liquid theory,  thus it is necessary to
analyze the validity of the  Hansen-Verlet freezing criterion in
this issue. Actually, the single order parameter
$C^{2D}_{sub}(G)=0.856$ is just relative to the 2D Hansen-Verlet
freezing criterion through the equation $S(k)=1/(1-C(k))$. For the
vortex system  in  layered superconductors with a large
anisotropy, the liquid freezes when the structure factor at $k=G$
reaches the value $6.94$. Interestingly, such a value is well
consistent with the value of S(k) at low-field regime along the
melting line obtained in previous work \cite{cor}.

Finally, we employ the freezing  criterion described above to locate
the melting lines. Critical fields for $LaO_{1-x}F_{x}FeAs$ $(x
=0.1, 0.075)$ at different temperatures are calculated, the melting
lines are shown in Fig. (3). For comparison, the melting line for
$BSCCO$ obtained in Ref. \cite{cor} is also given in the inset. One
can observe that  the monotonic meting lines for $LaOFeAs$ are
qualitatively alike but much lower than that for $BSCCO$. Recently,
a quite large vortex liquid  area in $LaO_{0.9}F_{0.1}FeAs$
compounds is observed in an experimental study based on the
measurement of Nernst signal \cite{zhu}, consistent with the melting
lines predicted  in the present work. Interestingly, we find that in
$LaOFeAs$ samples the melting line is changed substantially with the
variation of $F^{-}$ content, the melting line for
$LaO_{0.9}F_{0.1}FeAs$ is much higher than that for
$LaO_{0.925}F_{0.075}FeAs$.

The very recently discovered ternary  iron-arsenides
$(Ba,K)Fe_{2}S_{2}$ (122-type)\cite{Rotter} were suggested to be
nearly isotropic\cite{Yuan}. The present method to locate the
melting lines of course can not be applicable to these
superconductors with low anisotropy.

\section{Summary}

The methodological difference between the present work and the
previous study in Ref. \cite{col}  lies in the determination of 2D
OCP direct correlation functions.  In Ref. \cite{col} these
functions were obtained from Monte-Carlo simulations, while they are
achieved  $ab$ $inito$ from the microscopic vortex interaction in
the present study. Therefore, the present approach is more
intuitive, and the results are obtained without statistical errors
like in Monte-Carlo simulations. The method adopted in the present
work is also  different from  that in Ref. \cite{segupta} where the
density functional theory was first extended to  the flux-lattice
melting. A consistency to the Clausius-Clapeyron relation \cite{col}
can be reached with the use of the mean-field substrate model in the
present work, but can not be obtained in Ref. \cite{segupta}.

We use the density functional theory  and the  mean-field substrate
model to investigate the flux-lattice melting in iron-based
superconductors $LaO_{1-x}F_{x}FeAs (x=0.1, 0.075)$. The 2D OCP PDF,
2D OCP SSF and 2D  OCP DCF for $LaO_{0.9}F_{0.1}FeAs$ and
$LaO_{0.925}F_{0.075}FeAs$ are calculated at different temperatures
via the HNC closure and an efficient algorithm. It is shown that the
melting lines are quite low, well consistent  with an experimental
study on the Nernst signals \cite{zhu}. Furthermore, the melting
line is observed to change considerably with the modulation of
$F^{-}$ content in $F$-doped $LaOFeAs$ compounds, indicating that
the interaction between vortices crucially depends on the $F^{-}$
content. More experimental studies are needed to test the present
theoretical predications.

\section{Ackownledgements}
This work was supported by National Natural Science Foundation of
China under Grant Nos. 10574107 and 10774128, PCSIRT (Grant No.
IRT0754) in University in China,  National Basic Research Program of
China (Grant Nos. 2006CB601003 and 2009CB929104).

$^{\dag}$Corresponding author

\end{document}